\pdfoutput=1
\documentclass[cits]{PoS}
\usepackage{wrapfig}

\title{Brane tunneling and virtual brane-antibrane pairs}

\ShortTitle{Brane tunneling}

\author{\speaker{Adam R. Brown}%
         \thanks{This note is based on a recent paper \cite {bssw} (which exceeds the range of this note in including the effects of gravity, in presenting a fuller discussion of the UV physics, in considering cosmological consequences, and in providing more comprehensive references; this note exceeds the range of the paper in explicitly interpreting the novel features of brane tunneling as being due to virtual brane-antibrane pairs) and I thank my collaborators: Saswat Sarangi, Benjamin Shlaer and Amanda Weltman. I also thank Cliff Burgess, Dan Kabat, Philip Kim, Sarah Shandera, Erick Weinberg and my collaborators for invaluable feedback. Finally, I thank the organizers, Jean Orloff, G\'eraldine Servant and G\'erard Smadja, for the opportunity to participate. This material is based upon work supported under a National Science Foundation Graduate Research Fellowship.}\\
        Physics Department, Columbia University, New York, NY 10027, USA \\
        E-mail: \email{mr.adam.brown@gmail.com}}

\abstract{We survey barrier penetration by quantum tunneling for four cases: nonrelativistic point particles, scalar fields, relativistic point particles, and DBI branes. We examine two novel features that arise for DBI brane tunneling: the rate can sometimes increase as the barrier gets higher; and the instanton ``wrinkles''. We show that these features can be understood as the effect of the quantum sea of virtual brane-antibrane pairs. This sea exponentially augments the decay rate, with possible cosmological consequences.}

\FullConference{Carg\`ese Summer School: Cosmology and Particle Physics Beyond the Standard Models\\
                 July 30 - August 11, 2007\\
                 Institut d'\'Etudes Scientifiques de Carg\`ese}

\begin{document}

We will survey barrier penetration by quantum tunneling for four cases: nonrelativistic point particles, scalar fields, relativistic point particles, and DBI branes. We will discover two novel features that arise for DBI brane tunneling (and indeed already for relativistic point particles): the rate can sometimes increase as the barrier gets higher; and the instanton ``wrinkles''. We will show that these features can be understood as the effect of the quantum sea of virtual brane-antibrane pairs, and that this sea exponentially augments the decay rate. This work finds application in cosmology by providing an effective description of transitions in the landscape of string vacua \cite{bssw}.  We will see that DBI tunneling combines the complications of Coleman tunneling (see Fig. 1) and those of relativistic point particle tunneling, with no additional complications thrown in.

 \begin{figure}[h] 
   \centering
      \includegraphics[width=6in]{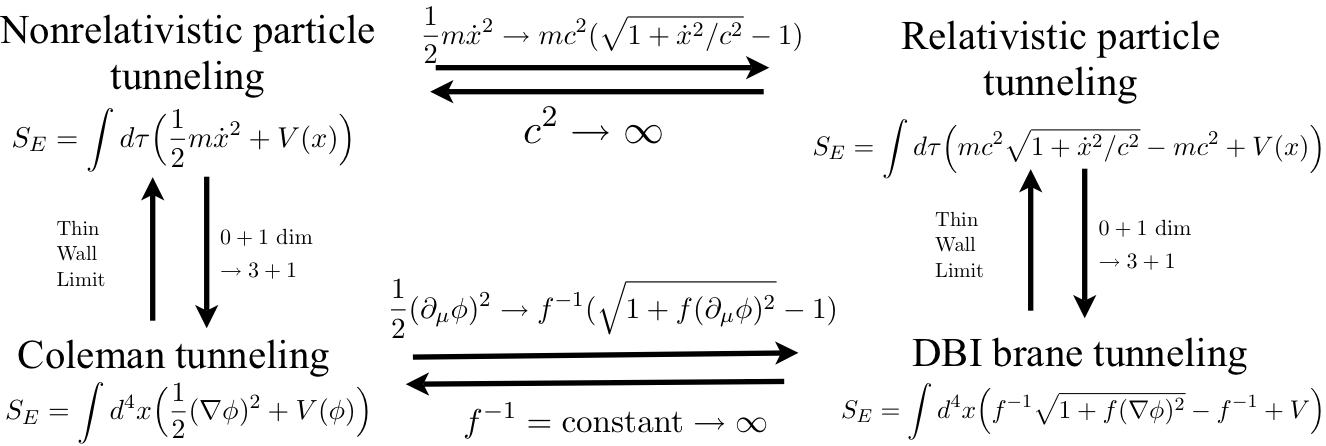} 
   \caption{The four tunneling cases we will discuss, and their relationships.}
   \label{four}
\end{figure}

\paragraph{Nonrelativistic point particle tunneling} \

\begin{wrapfigure}{r}{0.5\textwidth}
\centering
   \includegraphics[width=3in]{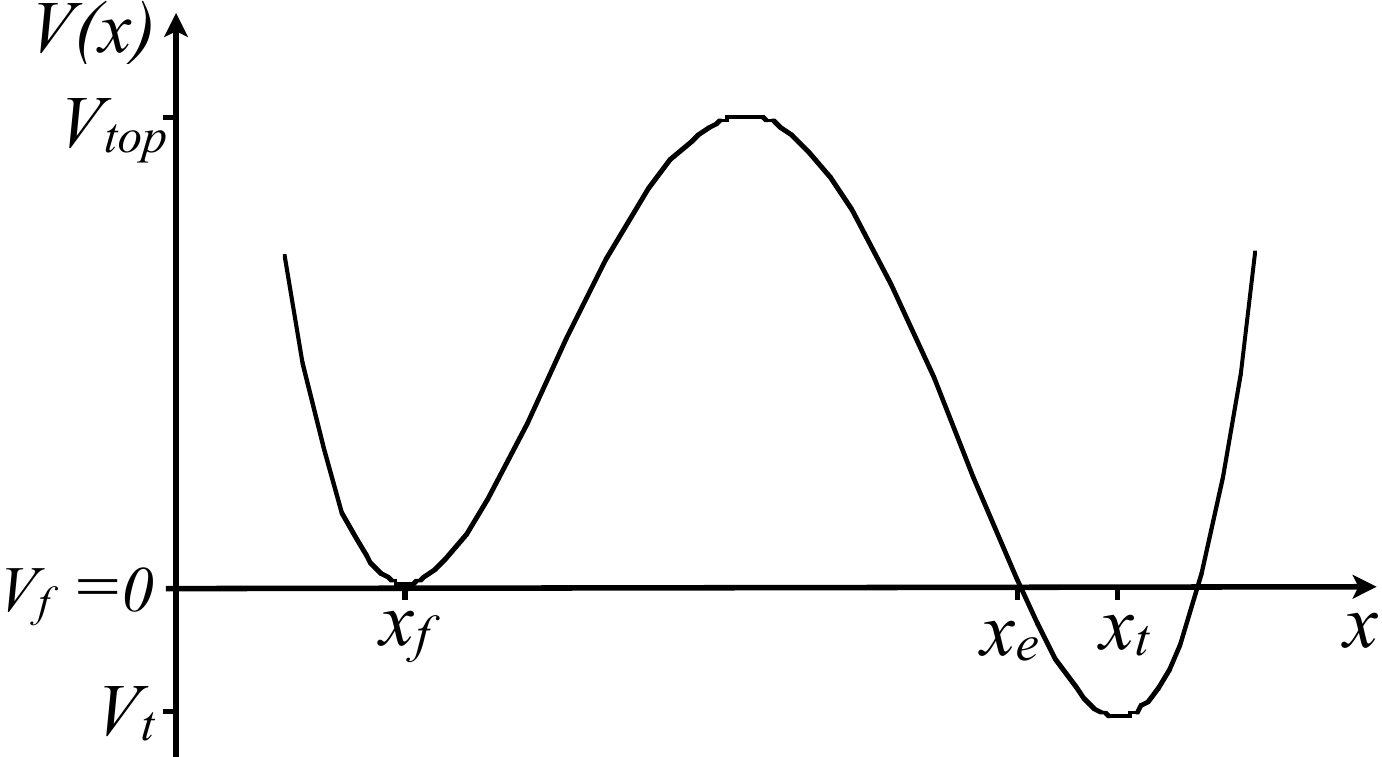} 
\centering
   \caption{A potential with a metastable minimum  (the ``false vacuum'' at $x_f$) and a stable minimum (the ``true vacuum'' at $x_t$). The escape point, $x_e$, has the same energy as the false vacuum. Since we are ignoring the effects of gravity throughout, we are free to set $V_{f}$ = 0. } 
\end{wrapfigure}

The tunneling rate of a nonrelativistic point particle is given by the Euclidean action of an instanton interpolating in Euclidean time from the false vacuum $x_{f}$ to the escape point $x_{e}$. In particular, $\Gamma \sim e^{-2 S_E}$ 
\begin{eqnarray}
S_{E} & = & \int d\tau \Bigl(\frac{1}{2}m \dot{x}^2 + V(x) \Bigl). \label{action1}
\end{eqnarray}

The most probable escape path - the instanton -  minimizes the Euclidean action and is found by solving the Euler-Lagrange equation and integrating.
\begin{eqnarray}
 \dot{x}^2 & = & 2 V(x)/m \label{eom1}\\
\Rightarrow \ S_{E} & = & \int_{x_f}^{x_e} dx \sqrt{2 m V(x)}
\end{eqnarray}
\begin{wrapfigure}{r}{.5\textwidth}
   \centering
   \includegraphics[width=0.492\textwidth]{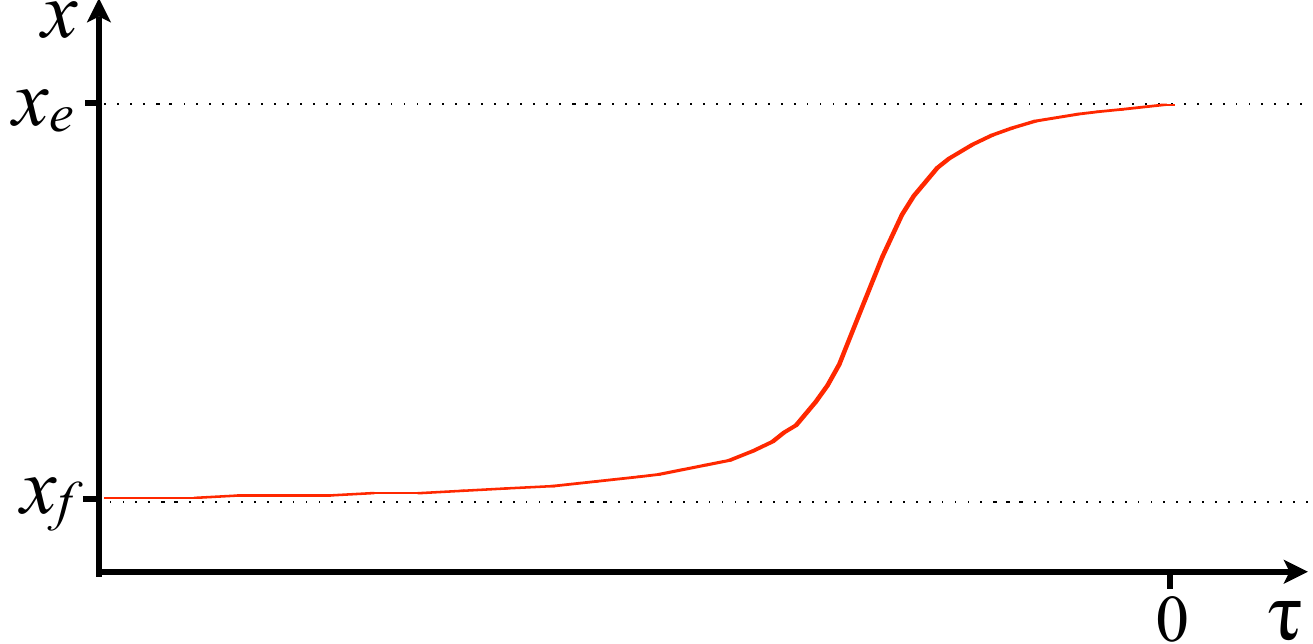}   
   \caption{The profile of the instanton, given by integrating Eq. 2. The instanton is the path, $x(\tau)$, in Euclidean time which takes you from $x_f$ (as $\tau \rightarrow -\infty$) to $x_e$ (conventionally at $\tau = 0$) with lowest Euclidean action. $\dot{x}$ is finite throughout.}
\end{wrapfigure}
This tunneling has two notable features, which are inherited by Coleman tunneling. 
\begin{enumerate}
\item The higher the barrier, the slower the tunneling. 
\item The Euclidean velocity $\dot{x}$ is kept finite throughout the instanton (witness Eq. 2) by the interplay, in Eq. \ref{action1}, of the slowing $\frac{1}{2} m \dot{x}^2$ term and the quickening $V(x)$ term. (See Fig.  3)
\end{enumerate}

\paragraph{Coleman tunneling} \

These calculations directly generalize, with one subtlety, to the tunneling of scalar fields in Minkowski space \cite{coleman}. The subtlety is that the analogue of $V(x)$ is not $V(\phi)$ but $U[\phi(\vec{x})] = \int d^3x \Bigl\{(\vec{\nabla} \phi(\vec{x}))^2 + V(\phi(\vec{x})) \Bigl\}$ so a scalar field cannot \emph{homogeneously} tunnel from $\phi_f$ to $\phi_e$: the Euclidean action would scale with the spatial volume. Instead the instanton describes the inhomogeneous nucleation of a ``bubble'' of approximately true vacuum separated from the false vacuum by a ``wall'' whose field value interpolates between them. The bubble must have zero energy ($U[\phi_{b}(\vec{x})] = 0$), and so must be large enough that the decrease in energy from its true-vacuum interior can compensate for the surface tension in the wall. The instanton has $O(4)$ symmetry, so in four dimensional spherical polar coordinates we have
\begin{equation}
S_{E}  =  2 \pi^2 \int d\rho \, \rho^3 \Bigl( \frac{1}{2} \dot{\phi}(\rho)^2 + V[\phi(\rho)] \Bigl). \label{action2}
\end{equation}
This differs from the point particle expression in two conventional ways ($x(\tau) \rightarrow \phi(\rho)$, the rescaling to eliminate the `$m$') and one substantive way (the appearance of the volume factor $\rho^3$). However, if the barrier is tall and thin, then the zero-energy bubble is much larger than its wall is thick (the ``thin wall'' regime), so the volume factor is effectively constant throughout the wall. We can then solve analytically for the instanton shape and the decay rate $\Gamma \sim \textrm{exp}[{-\frac{27 \pi^2 S_1^4}{2 (V_f - V_t)^3}}]$, with
\begin{eqnarray}
\dot{\phi}^2 & = & 2 \, V(\phi) \\
S_1 & = & \int d\phi \sqrt{2 V(\phi)}. 
\end{eqnarray}

\paragraph{Relativistic point particle tunneling} \

For relativistic electrons in an electrostatic potential $V$, the tunneling rate (which may soon be experimentally measured in graphene \cite{graphene}) is given by the Euclidean action of an instanton, $\Gamma \sim e^{-2S_{E}}$, which we compute by solving the Euler-Lagrange equations and integrating
\begin{eqnarray}
S_{E} & = & \int d\tau \Bigl(m c^2 \sqrt{1 + \dot{x}^2/c^2} -mc^2 + V(x) \Bigl) \label{action3} \\
\Rightarrow \ \ \ \ \dot{x}^2 & = & \frac{V(x) (2m - V(x)/c^2)}{(m -V(x)/c^2)^2} \label{eq:ep2} \\ 
\Rightarrow \ \ S_{E} & = & \int_{x_f}^{x_e} dx \sqrt{V(x)(2m - V(x)/ c^2)}.
\end{eqnarray}
Eq. \ref{action3} can be brought into standard form by redefining $V(x) \rightarrow V(x) + mc^2$, but our form makes it clear that for small barriers ($V \ll m c^2$) we recover the nonrelativistic results. However, for tall barriers we have two novel features.
\begin{enumerate}
\item For $V(x)>mc^2$, tunneling gets faster with increasing barrier height. 
\item As $V(x) \rightarrow mc^2$, $\dot{x}$ becomes infinite (the $V(x)$ term in Eq. \ref{action3} overwhelms the $mc^2\sqrt{1+\dot{x}^2/c^2}$).
\end{enumerate}

\begin{wrapfigure}{r}{.5\textwidth}   \centering
   \centering
   \includegraphics[width=.5\textwidth]{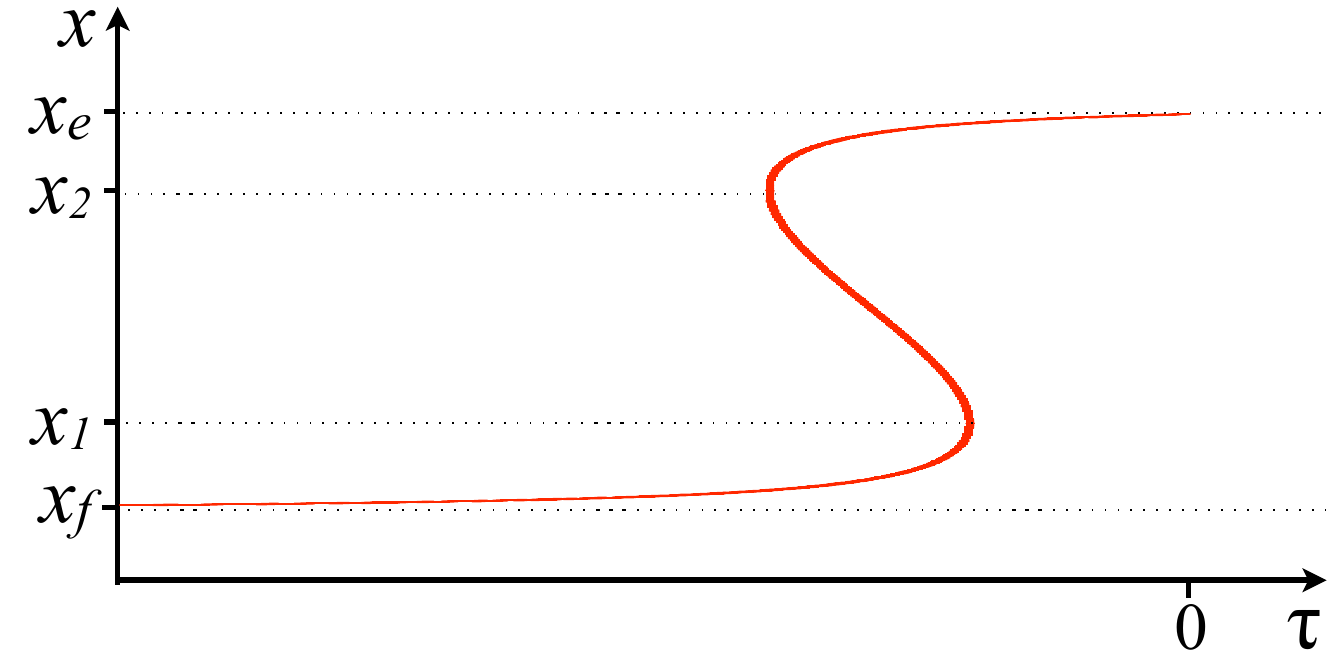} \label{wrinkle}
   \caption{The profile of the instanton, from integrating Eq.  8. Unlike the non-relativistic case (see Fig. 3), the tunneling electron apparently goes backwards in Euclidean time (the instanton ``wrinkles''). We interpret the segment of the instanton where this happens (i.e. between $x_1$ \& $x_2 \rightarrow V(x)>mc^2$) as a positron going forward in time.} 
\end{wrapfigure}

\begin{figure}[h] 
   \centering
   \includegraphics[width=3.5in]{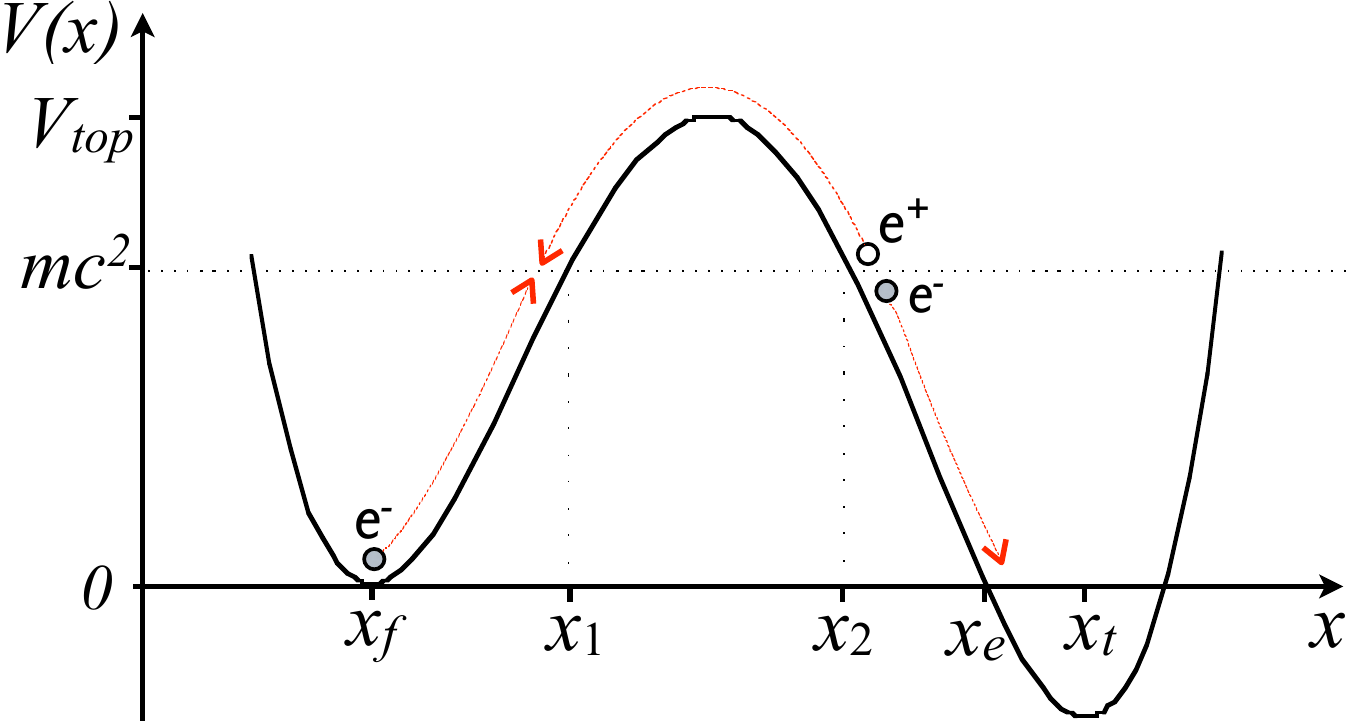} 
   \caption{Positron-assisted tunneling. The electron at $x_f$ starts to tunnel up the barrier. A virtual pair is created at $x_2$, and the positron tunnels to $x_1$ and annihilates the initial electron. Meanwhile, the electron created at $x_2$ tunnels to $x_e$ where it appears as the physical electron.}
   \label{apat}
\end{figure}

We can interpret this behaviour in terms of positron-assisted tunneling. For sections of the potential where $V(x) > mc^2$, rather than an electron tunneling \emph{forward} across the barrier, it is favourable to create a virtual pair on the true vacuum side of the barrier and have the positron tunnel \emph{backward} (see Fig. \ref{apat}). 

To see that this is correct, consider the contribution to the Euclidean action (Eq. \ref{action3}) incurred crossing a segment of the barrier $x$ to $x + dx$. To traverse the segment with an electron, the effective potential is $V(x)$; to traverse with a (oppositely charged) positron, the effective potential is $-V(x)$, plus $2mc^2$ from having to pair create. Positrons are favoured exactly when $V(x) > mc^2$. 

Seen in this context the two novel features of relativistic tunneling are easy to understand. 
\begin{enumerate}
\item Those parts of the barrier with $V(x) > mc^2$ are traversed not by an electron, but by a positron. 
 Positrons, being oppositely charged, actually prefer higher $V(x)$.
\item The ``wrinkling'' (see Fig. 4) of the instanton is nothing more than the familiar fact that a positron may be interpreted as an electron going backwards in time. 
\end{enumerate}

We then have three regimes (with a distracting subtlety when $V_{t} < V_{top} - 2mc^2 < V_{f}$): 

\begin{displaymath}
\begin{array}{lllllll}
& & V_{top} - V_{f} & < & mc^2 & \rightarrow & \textrm{electron tunneling} \\
mc^2 & < & V_{top} - V_{f} & < & 2mc^2 & \rightarrow & \textrm{electron tunneling assisted by \emph{virtual} electron-positron pair.} \\
2mc^2 & < & V_{top} - V_{f} && & \rightarrow & \textrm{Schwinger production of \emph{real} electron-positron pairs. Whenever} \\
& & & & & & \textrm{this can happen, it dominates (and renders meaningless) tunneling.}
\end{array}
\end{displaymath}

\paragraph{Brane tunneling} \

We combine the lessons of the last two sections to discuss the tunneling of DBI branes.
\begin{equation}
S_{Euclidean \, \, DBI} = 2\pi^2  \int d\rho \, \rho^3  \Bigl( f(\phi)^{-1} \sqrt{1 + f(\phi)\dot{\phi}^2} - f(\phi)^{-1}  +  V(\phi) \Bigl)
\end{equation}
Here $\phi(x)$ is a four dimensional field providing an effective description of the embedding of the brane in a higher dimensional target space. $f(\phi)^{-1}$ is a warp factor that can depend on $\phi$ - though the novel features we will discuss appear already for constant $f^{-1}$. 

\begin{wrapfigure}{r}{0.5\textwidth}
   \centering
   \includegraphics[width=0.5\textwidth]{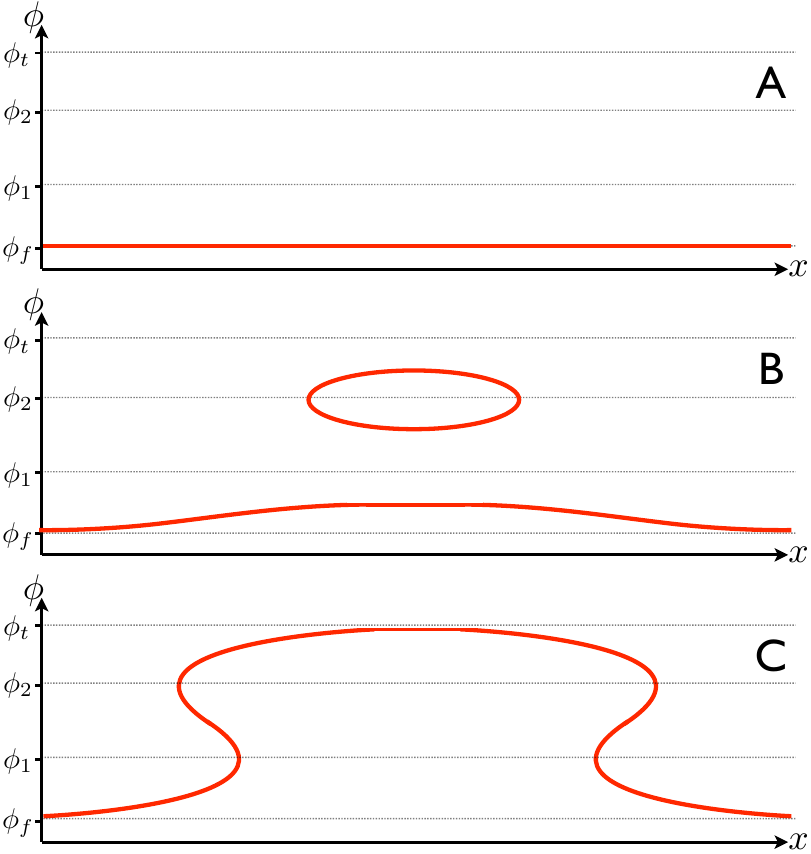} 
   \caption{A sequence of spatial slices through the thin-wall antibrane-assisted bubble nucleation instanton (compare to spatial slices through Fig. 4). (a) The brane is in the false vacuum. (b) A pair is created at $\phi_2$. (c) The antibrane annihilates part of the original brane (at $\phi_1$), leaving a brane with a zero-energy bubble of true vacuum; the bubble then classically expands.}
\end{wrapfigure}

Just as for scalar field tunneling, the false vacuum decays by nucleating bubbles of true vacuum, and in the thin wall regime we can give analytic expressions for the shape of the instanton
\begin{equation}
\dot{\phi}^2   =  \frac{V (2 - f V)}{(1 - f V)^2},
\end{equation}
and for the decay rate $\Gamma \sim \textrm{exp}[{-\frac{27 \pi^2 S_1^4}{2 (V_f - V_t)^3}}]$, where
\begin{equation}
S_{1}  =  \int d\phi \sqrt{V(\phi) (2 - f(\phi) V(\phi))}.
\end{equation}
Just as for electron tunneling, however, we recover the two novel features related to virtual pair creation. For $V_{top} > f^{-1}$ the optimal decay route makes use of a virtual brane-antibrane pair. For $V_{top} > 2f^{-1}$ a real brane-antibrane pair is created. Notice that the brane pairs, real or virtual, are not created homogeneously - the Euclidean action of such a creation would scale with the spatial volume. Instead a spatial cross section through the pair gives two coincident three-balls, joined at the bounding two-sphere. Comparing Eqs. 6 \& 12, brane decay is always faster than Coleman decay in the same potential, with the deviation vanishing for small barriers ($V \ll f^{-1}$).  The quantum sea of virtual brane-antibrane pairs exponentially augments the decay rate.

\end{document}